\documentclass[
  aip,
  cha,
  amsmath,amssymb,
  reprint,
]{revtex4-1}

\usepackage{graphicx}
\usepackage{dcolumn}
\usepackage{bm}
\usepackage{hyperref}
\usepackage[utf8]{inputenc}
\usepackage[T1]{fontenc}
\usepackage{mathptmx}
\usepackage{url}
\usepackage{multirow}
\usepackage{subfig}
\usepackage[bottom]{footmisc}
\usepackage{siunitx}
\usepackage{etoolbox}
\makeatletter
\def\@email#1#2{%
 \endgroup
 \patchcmd{\titleblock@produce}
  {\frontmatter@RRAPformat}
  {\frontmatter@RRAPformat{\produce@RRAP{*#1\href{mailto:#2}{#2}}}\frontmatter@RRAPformat}
  {}{}
}%
\makeatother
\begin{document}

\preprint{AIP/123-QED}

\title[Kolmogorov-Arnold Networks for Turbulence Anisotropy Mapping]{Kolmogorov-Arnold Networks for Turbulence Anisotropy Mapping}

\author{N. Kalia}
\email{nkalia@uwaterloo.ca}
\affiliation{Department of Mechanical and Mechatronics Engineering, University of Waterloo}

\author{R. McConkey}
\affiliation{Department of Mechanical Engineering, Massachusetts Institute of Technology}

\author{E. Yee}
\affiliation{Department of Mechanical and Mechatronics Engineering, University of Waterloo}

\author{F.S. Lien}
\affiliation{Department of Mechanical and Mechatronics Engineering, University of Waterloo}

\date{\today}

\begin{abstract}
This study evaluates the generalization performance and representation efficiency (parsimony) of a previously introduced Tensor Basis Kolmogorov–Arnold Network (TBKAN) architecture for data-driven turbulence modeling. The TBKAN framework replaces the multi-layer perceptron (MLP) used in either the standard or modified Tensor Basis Neural Network (TBNN) with a Kolmogorov-Arnold network (KAN), which significantly reduces the model complexity while providing a structure that potentially can be used with symbolic regression to provide a physical interpretability that is not available in a ``black box'' MLP. While some prior work demonstrated TBKAN’s feasibility for modeling a ``simple'' flat plate boundary layer flow, this study extends the TBKAN architecture to model more complex benchmark flows---in particular, square duct and periodic hills flows which exhibit strong turbulence anisotropy, secondary motion, and flow separation and reattachment. A realizability-informed loss function is employed to constrain the model predictions, and, for the first time, TBKAN predictions are stably injected into the Reynolds-averaged Navier–Stokes equations to provide {\it a posteriori\/} predictions of the mean velocity field. 
Results show that a TBKAN achieves comparable or a slightly improved accuracy relative to a TBNN (based on a MLP), while using significantly required fewer parameters to achieve this performance in comparison to a TBNN---both TBNN and TBKAN successfully captures key flow features in the square duct and periodic hills flows (e.g., secondary motions of the second kind, separation and reattachement zones, etc.) and demonstrate a significantly improved predictive performance relative to the conventional $k$-$\omega$ SST turbulence closure model. Finally, this study highlights the modular and interpretable structure of the TBKAN which can potentially facilitate symbolic regression for the derivation of algebraic stress models from data, providing an attractive alternative for turbulence closure modeling in complex flow regimes.
\end{abstract}

\maketitle
\section{\label{sec:introduction}Introduction}

Turbulence modeling is a critical challenge in computational fluid dynamics (CFD), particularly for engineering applications where accurate predictions of flow behavior are essential. High-fidelity approaches such as direct numerical simulation (DNS) and large-eddy simulation (LES) provide detailed insights into turbulent flows, but their computational costs remain high~\cite{moin1998direct, sagaut2006large}. DNS resolves all relevant turbulence scales down to the Kolmogorov length scale, where viscous dissipation dominates. While this makes DNS the most accurate method, its computational cost increases rapidly with Reynolds number, making it impractical for most real-world applications. LES reduces this cost by resolving the larger turbulent structures while modeling the small-scale effects, typically capturing turbulence down to the Taylor microscale. However, in wall-bounded flows, LES still requires a fine near-wall resolution, significantly increasing computational costs. Given these constraints, Reynolds-averaged Navier-Stokes (RANS) models continue to be the most widely used approach in industrial and engineering simulations, providing a balance between accuracy and computational efficiency~\cite{duraisamy2019}.

RANS models rely on turbulence closures to approximate the effects of unresolved scales, typically through eddy-viscosity models that relate the Reynolds stress tensor to the mean strain rate tensor. The Boussinesq hypothesis, which serves as the foundation for these models, assumes that the anisotropy of the Reynolds stress tensor can be expressed as a function of the mean strain rate. This assumption simplifies turbulence modeling but inherently limits its ability to capture complex flow phenomena such as turbulence-driven secondary flows, streamline curvature effects, and strong anisotropy in turbulence structures. Traditional two-equation models, such as the $k$-$\varepsilon$ and $k$-$\omega$ shear stress transport (SST) models, perform well in simple flow configurations but frequently struggle in cases where turbulence anisotropy plays a significant role. This limitation arises because these models assume implicitly that turbulence behaves isotropically with respect to the mean strain rate, which is often not the case in real-world engineering applications.

An alternative to eddy-viscosity models is the Reynolds stress transport model (RSM), introduced by~\citet{launder1975progress}, which provides a more detailed representation of turbulence by directly solving the transport equations for each component of the Reynolds stress tensor. This approach avoids the isotropy assumption but introduces additional computational complexity, requiring closure models for the pressure-strain correlation, turbulent diffusion, and dissipation terms. While RSMs improve predictive accuracy for highly anisotropic flows, they are significantly more expensive computationally, making them less attractive for industrial applications where cost-effective turbulence modeling is crucial.

Due to these inherent limitations, approaches based on machine learning have been explored as a means to enhance RANS turbulence closures~\cite{duraisamy2015data, tracey2015machine, singh2017machine, parish2017paradigm}. By leveraging high-fidelity data from DNS and LES, data-driven models can either provide corrections to predictions of the Reynolds stress fields provided by RANS or directly model the anisotropy tensor. One of the most significant breakthroughs in this area was introduced by~\citet{Ling2016}, who proposed the Tensor Basis Neural Network (TBNN). This model marked a major advancement by embedding Galilean and rotational invariance into a neural network architecture, ensuring that predictions of the Reynolds stress anisotropy tensor adhered to fundamental physical constraints. The TBNN architecture expresses the anisotropy tensor as a linear combination of tensor basis functions, with the learned coefficients capturing the complex turbulence interactions. By explicitly incorporating these symmetry constraints, these models for the anisotropy tensor significantly improved predictive accuracy compared to traditional eddy-viscosity models, particularly in flows characterized by strong turbulence anisotropy.

Subsequent investigations have extended the TBNN framework. For example,~\citet{mcconkey2024realizability} introduced realizability constraints to improve the numerical stability and generalization of the anisotropy predictions in RANS solvers. Others have explored alternative regression methods while retaining tensor basis structures, such as Kaandorp and Dwight’s~\cite{kaandorp2020data} tensor-basis random forest model (TBRF). This research demonstrated that ensemble-based approaches could offer comparable accuracy with greater training stability.~\citet{wang2017physics} proposed a physics-informed machine learning approach using random forests to model the Reynolds stresses for RANS turbulence closure, emphasizing generalization across a range of Reynolds numbers and flow geometries.

An alternative methodology referred to as the Field Inversion and Machine Learning (FIML) framework has been developed by~\citet{singh2017machine} and~\citet{parish2017paradigm}. These researchers proposed a hybrid modeling strategy in which model-form discrepancies are first inferred using an inverse analysis and then approximated subsequently using supervised learning techniques. These approaches often retain the structure of traditional turbulence closure models (e.g., Spalart-Allmaras or $k$-$\omega$ models) but enhance their predictive capacity with learned corrections trained on experimental and/or high-fidelity data from numerical simuations. The FIML methodology has demonstrated notable success in separated flows and airfoil applications~\cite{singh2017machine}. Recent work by ~\citet{kaszuba2025implicit} proposed an alternative to symbolic tensor basis engineering by leveraging equivariant Euclidean neural networks (e3nn) to model turbulence closure terms. Their architecture inherently embeds symmetry properties and was shown to implicitly recover Pope's tensor basis formulation as a special case. 

Despite the success of these methods, most machine learning-based turbulence models, including TBNN, rely on multi-layer perceptrons (MLPs) as universal function approximators~\cite{Ling2016, wang2017physics, mcconkey2022deep}. MLPs are powerful tools in deep learning, offering the ability to approximate complex, nonlinear functions through a hierarchical network of neurons. Their universal approximation capability makes them well-suited for learning intricate relationships within turbulence data. However, their effectiveness in turbulence modeling depends heavily on network architecture, dataset availability, and training strategies. In practice, achieving optimal performance often requires careful tuning of hyperparameters and managing large numbers of trainable parameters. Additionally, while MLPs are highly flexible, they do not inherently encode physical constraints, which can sometimes lead to difficulties in interpretability and generalization, particularly when applied to flow conditions beyond the data used in their training. These considerations motivate the exploration of alternative neural network architectures that can maintain the expressive power of MLPs while incorporating structured representations that may enhance efficiency, interpretability, and robustness in turbulence modeling. 

Kolmogorov-Arnold networks (KANs) proposed by~\citet{liu2024kan} offer an alternative approach to function approximation by leveraging the Kolmogorov-Arnold representation theorem, which states that any continuous function can be expressed as a finite sum of univariate functions applied to linear combinations of input variables~\cite{liu2024kan}. Unlike MLPs, which rely on multiple hidden layers to learn complex transformations, KANs explicitly construct structured functional decompositions. By structuring the functional representation explicitly, KANs can reduce the number of parameters needed without sacrificing flexibility, offering a more interpretable alternative to traditional deep neural networks. Because of these features, KANs can be especially useful in turbulence modeling, where having more control and interpretability is important. A recent study by~\citet{mcconkey2024realizability} explored a preliminary integration of KANs into turbulence modeling, highlighting their potential for improving generalization and realizability when predicting Reynolds stress anisotropy. In this study, the researchers applied KANs within a modified TBNN framework to assess their effectiveness in comparison to MLP-based models, investigating their ability to generalize and improve realizability in machine learning-based turbulence modeling for a ``simple'' test case---namely, a turbulent flat plate boundary-layer flow.

The primary objective of this study is to assess the feasibility of using KANs as an alternative universal function approximator for machine learning-based turbulence closure models. To achieve this, KANs will be used within the modified TBNN framework and evaluated against conventional turbulence closure models across two benchmark cases: namely, the square duct and the periodic hills flow cases. 
This paper extends a preliminary investigation~\cite{mcconkey2024realizability} of the utility of KANS for modeling the flow over a flat-plate boundary layer to more complex flows, accompanied with a more systematic comparison of the results to MLP-based models. The square duct case provides a challenging scenario where secondary flows emerge due to turbulence-driven anisotropy, which eddy-viscosity models typically fail to predict accurately~\cite{kaandorp2020data}. The periodic hills case introduces adverse pressure gradients and separation effects, requiring an accurate representation of the Reynolds stress anisotropy to capture reattachment dynamics~\cite{wang2017physics}. By evaluating the performance of KAN-based models in these diverse flow configurations, the objective of this study is to determine whether KANs provide improved realizability, accuracy, and generalizability compared to the more conventional MLP-based approaches.

\section{\label{sec:theory}Kolmogorov-Arnold Networks (KANs)}

The Kolmogorov-Arnold network, proposed by~\citet{liu2024kan}, is based on the Kolmogorov-Arnold representation theorem. This network is introduced as a promising alternative to a MLP. Neural networks, and in particular multi-layer perceptrons, have served as the foundational architecture for a wide range of applications in machine learning. Since Rosenblatt's original introduction of the perceptron \cite{rosenblatt1958perceptron}, MLPs have evolved into deep neural networks capable of approximating highly nonlinear functions. MLPs have been widely used in turbulence modeling and are often the ``tool'' of choice owing to their flexibility and proven capability as universal function approximators. Their ability to model complex nonlinear relationships has made them a go-to architecture in many data-driven closure modeling frameworks. However, MLPs are inherently unstructured, and their dense parameterization can make interpretation challenging. This study explores whether an alternative architecture based on KANs can serve as a viable replacement within the turbulence modeling context. By leveraging structured, B-spline-based function representations, KANs may offer advantages in terms of interpretability, parameter efficiency, and physical consistency, particularly in tasks where realizability and function smoothness are important. Whether these benefits translate into improved predictive performance in Reynolds stress modeling is a key question addressed in this work.

\subsection{Theoretical background}

Kolmogorov-Arnold networks, which were recently proposed by ~\citet{liu2024kan}, are inspired by the Kolmogorov-Arnold representation theorem, which provides a constructive framework for approximating multivariate continuous functions using compositions of univariate functions. This theorem states that any continuous function \( f: \mathbb{R}^n \to \mathbb{R} \) can be expressed as a finite sum of continuous univariate functions composed with linear combinations of the inputs. Formally, there exist continuous functions \( \phi_i \) and \( \psi_{ij} \) such that:

\begin{equation}
f(x_1, x_2, \ldots, x_n) = \sum_{i=1}^{2n+1} \phi_i\left( \sum_{j=1}^n \psi_{ij}(x_j) \right)\ ,
\label{eq:kan_theorem}
\end{equation}
where each \( \psi_{ij} \) acts on a single input dimension \( x_j \), and each \( \phi_i \) acts on a linear combination of these transformed inputs. This formulation highlights the potential for structured function approximation through composition and summation of simpler, interpretable components.

KANs leverage the Kolmogorov-Arnold representation theorem to provide an alternative representation based on compositions of learnable univariate functions to that given by affine transformations followed by the pointwise action of nonlinear activation functions used in MLPs. In contrast, learnable univariate functions used in KANS are applied to linear combinations of the input features, enabling the model to approximate nonlinear relationships in a more explicitly compositional form. The univariate functions are typically implemented using splines, such as B-splines, which are defined over a fixed grid and trained by adjusting their control point values. This construction offers a flexible and localized way to represent nonlinear mappings while preserving continuity and smoothness of the mapping.

The resulting architecture provides an alternative to MLPs. While MLPs model nonlinearity through the imposition of an activation function that is applied after an affine transformation of the inputs to a particular network layer, KANs embed the nonlinearity directly into learnable activation functions that reside on the edges of the network between the nodes. This structure changes how the model captures nonlinearity, making it more focused on local patterns and smooth transitions. This is useful in cases such as turbulence modeling, where we are concerned about interpretability and avoidance of noisy predictions.

In this work, KANs are explored as the function approximation module within a turbulence modeling framework. Their structured form aligns well with scientific machine learning objectives and provides a promising avenue for learning mappings in a data-driven approach for the construction of a turbulence closure model. The next subsection describes how this architecture is constructed and implemented in practice.

\subsection{KAN architecture and implementation}

Kolmogorov-Arnold networks differ fundamentally from traditional neural networks in their internal structure and approach to function approximation. Rather than using fully-connected layers composed of affine transformations followed by the application of nonlinear activations as in MLPs, KANs are constructed using layers where each edge linking the nodes of the network applies a learnable univariate function (represented using a B-spline) to a linear combination of its inputs. This design follows directly (is inspired) from the Kolmogorov-Arnold representation theorem, which states that a multivariate function can be represented as a sum of compositions of linear combinations of univariate functions. The architecture introduced by~\citet{liu2024kan} enforces this idea through the use of B-spline parameterizations.

More specifically, each learnable function in KAN is represented by a B-spline that is a piecewise polynomial function defined over a fixed grid of input values. These spline functions are initialized to smooth, low-order polynomials (typically cubic) and are updated during training to best fit the target mapping. The locality of B-splines ensures that only a few control points influence the output for any given input, enhancing both interpretability and numerical stability. The primary architectural hyperparameters that control the complexity and expressiveness of a KAN are the grid size (i.e., the number of control points), the spline order, and the width, which collectively determine how many input features are used per node.

The output from a single KAN node is computed by applying a spline function to a weighted sum of inputs, so
\begin{equation}
z = \phi\left(\sum_{j=1}^{w} \omega_j x_j\right)\ ,
\end{equation}
where \( \omega_j \) are the learnable weights, \( x_j \) are the selected input features, and \( \phi(\cdot) \) is a trainable univariate spline function. Unlike the fully connected layers and activation functions used in MLPs, KANs apply localized spline functions after a sparse linear projection. This structure can still capture complex nonlinear behavior while using fewer parameters, which often allows for shallower networks without sacrificing expressiveness.

In this study, we utilize the open-source pykan library (\url{https://github.com/KindXiaoming/pykan}) provided by~\citet{liu2024kan} as the implementation of KAN for turbulence closure modeling. This Python-based framework is built on PyTorch and supports the construction of differentiable spline functions with customizable grid sizes and orders. We integrated this implementation directly into our Tensor Basis framework, replacing the MLP layers used to learn the tensor basis coefficients with KAN layers. The overall network architecture and training procedure are described in the next section.

\subsection{Modified TBNN framework (TBKAN)}

This study extends significantly some recent work conducted by \citet{mcconkey2024realizability} through the incorporation of Kolmogorov-Arnold networks into the modified TBNN framework---replacing, as such, the MLP in this framework with the KAN to provide a modified TBNN framework based on KAN (viz., TBKAN framework). It should be noted that \citet{mcconkey2024realizability} already applied the TBKAN architecture to the flat plate boundary layer flow case. While the MLP-based modified TBNN achieved slightly lower prediction errors in this case, the KAN-based model was shown to produce comparable results using a significantly simpler architecture with a demonstrated improved physical realizability of the predicted anisotropy tensors. Encouraged by these findings, the present work extends the TBKAN formulation to more complex turbulent flows: the square duct and periodic hills flow cases. These benchmark configurations are particularly challenging due to the presence of secondary flows and flow separation, where anisotropy effects dominate and traditional RANS closures often fail. In this study, we retain the modified TBNN architecture introduced by~\citet{mcconkey2024realizability}, in which the Reynolds stress anisotropy tensor is expressed as a linear combination of invariant tensor bases~\cite{Pope1975}, but replace the MLP used to learn the basis coefficients with a structured, B-spline-based KAN. The goal is to evaluate whether the benefits of TBKAN (e.g., interpretability) can be retained in more anisotropic and nonlinear flow scenarios. The following section outlines the architecture, training methods, and evaluation metrics used for this extended investigation.

\section{\label{sec:methodology}Methodology}

The overall methodology employed in this study follows the framework introduced by~\citet{mcconkey2024realizability}, where a modified TBNN is trained to predict the anisotropy tensor based on various scalar invariants, including those derived from the mean strain-rate and rotation-rate tensors. The key novelty in the present work lies in the replacement of the MLP used to learn the tensor basis coefficients with a KAN, resulting in the Tensor Basis Kolmogorov-Arnold network (or, TBKAN) architecture. All other components of the modeling approach, including the input feature construction, tensor basis formulation, imposition of physical realizability constraints, and formuation of the loss function remain consistent with the original formulation of the modified TBNN framework.

For completeness, the key elements of the framework are briefly summarized in the following subsections. 

\subsection{Reynolds-averaged Navier–Stokes and closure modeling}

The time-averaged, incompressible steady-state Reynolds-averaged Navier–Stokes equations for a Newtonian fluid assume the following form:
\begin{equation}
U_i \frac{\partial U_j}{\partial x_i} = -\frac{1}{\rho} \frac{\partial P}{\partial x_j} + \nu \frac{\partial^2 U_j}{\partial x_i \partial x_i} - \frac{\partial \tau_{ij}}{\partial x_i}\ ,
\label{eq:rans_momentum}
\end{equation}
where \( U_j \) ($j=1,2,3$) is the mean velocity, \( P \) is the mean pressure, \( \nu \) is the kinematic viscosity, \( \rho \) is the fluid density, and \( \tau_{ij} \equiv -\overline{u_i' u_j'} \) is the Reynolds stress tensor (where $u_i'$ is the $i$-th component of the velocity fluctuation---deviation of the instantaneous velocity from the mean velocity). Equation~\eqref{eq:rans_momentum} is unclosed due to the appearance of \( \tau_{ij} \), which must be modeled to enable a numerical solution.

A conventional form of the Reynolds stress tensor involves a decomposition into its isotropic and anisotropic components to give
\begin{equation}
\tau_{ij} = \frac{2}{3}k \delta_{ij} + a_{ij}\ ,
\end{equation}
where \( k \equiv \frac{1}{2} \tau_{kk} \) is the turbulent kinetic energy and \( a_{ij} \) is the anisotropy tensor. The isotropic part is typically absorbed into a modified pressure term in Eq.~\eqref{eq:rans_momentum} and the governing equations are solved in terms of \( a_{ij} \).

Traditional linear eddy-viscosity models (LEVMs) model \( a_{ij} \) using the Boussinesq approximation, which relates the anisotropy tensor to the mean strain-rate tensor \( S_{ij} \) as follows:
\begin{equation}
a_{ij} = -2 \nu_t S_{ij}\ ,
\end{equation}
where
\begin{equation}
S_{ij} \equiv \frac{1}{2} \left( \frac{\partial U_i}{\partial x_j} + \frac{\partial U_j}{\partial x_i} \right)\ ,
\end{equation}
is the mean-strain rate tensor and \( \nu_t \) is the eddy viscosity. This formulation assumes that the turbulent stresses are aligned with \( S_{ij} \), that the closure is isotropic, and that it depends only on local gradients of the mean velocity. While computationally convenient and numerically stable due to its implicit implementation, LEVMs often fail to capture essential features such as secondary flows, streamline curvature effects, and anisotropy in wall-bounded turbulence.

To address these limitations,~\citet{Pope1975} proposed a more general nonlinear eddy-viscosity framework based on a tensor basis expansion. Using the Cayley–Hamilton theorem, the anisotropy tensor is expressed as
\begin{equation}
a_{ij} = 2k \sum_{n=1}^{10} g_n \hat{T}^{(n)}_{ij}\ ,
\end{equation}
where \( \hat{T}^{(n)}_{ij} \) are ten (non-dimensionalized) invariant tensor basis functions derived from the normalized mean strain-rate and mean rotation-rate tensors, and \( g_n \) are scalar coefficients that depend on the invariants of these tensors as well as other scalar invariants that characterize the flow. This formulation preserves Galilean invariance and allows for the modeling of complex anisotropic turbulence behavior. ~\citet{Ling2016} leveraged this formulation in their TBNN formulation, where the coefficients \( g_n \) are learned from data using a neural network trained on DNS-labeled anisotropy tensor data.

The ten basis tensors \( \hat{T}^{(n)}_{ij} \), constructed from the normalized strain-rate tensor \( \hat{S}_{ij} \) and rotation-rate tensor \( \hat{R}_{ij} \), are defined as follows (see~\citet{Pope1975}:
\begin{align*}
    \hat{T}^{(1)}_{ij} &= \hat{S}_{ij}\ , \nonumber \\
    \hat{T}^{(2)}_{ij} &= \hat{S}_{ik} \hat{R}_{kj} - \hat{R}_{ik} \hat{S}_{kj}, \nonumber \\
    \hat{T}^{(3)}_{ij} &= \hat{S}_{ik} \hat{S}_{kj} - \tfrac{1}{3} \hat{S}_{kl} \hat{S}_{lk} \delta_{ij}\ , \nonumber \\
    \hat{T}^{(4)}_{ij} &= \hat{R}_{ik} \hat{R}_{kj} - \tfrac{1}{3} \hat{R}_{kl} \hat{R}_{lk} \delta_{ij}\ , \nonumber \\
    \hat{T}^{(5)}_{ij} &= \hat{R}_{ik} \hat{S}_{kl} \hat{S}_{lj} - \hat{S}_{ik} \hat{S}_{kl} \hat{R}_{lj}\ , \nonumber \\
    \hat{T}^{(6)}_{ij} &= \hat{R}_{ik} \hat{R}_{kl} \hat{S}_{lj} + \hat{S}_{ik} \hat{R}_{kl} \hat{R}_{lj} - \tfrac{2}{3} \hat{S}_{kl} \hat{R}_{lm} \hat{R}_{mk} \delta_{ij}\ , \nonumber \\
    \hat{T}^{(7)}_{ij} &= \hat{R}_{ik} \hat{S}_{kl} \hat{R}_{lm} \hat{R}_{mj} - \hat{R}_{ik} \hat{R}_{kl} \hat{S}_{lm} \hat{R}_{mj}\ , \nonumber \\
    \hat{T}^{(8)}_{ij} &= \hat{S}_{ik} \hat{R}_{kl} \hat{S}_{lm} \hat{S}_{mj} - \hat{S}_{ik} \hat{S}_{kl} \hat{R}_{lm} \hat{S}_{mj}\ , \nonumber \\
    \hat{T}^{(9)}_{ij} &= \hat{R}_{ik} \hat{R}_{kl} \hat{S}_{lm} \hat{S}_{mj} + \hat{S}_{ik} \hat{S}_{kl} \hat{R}_{lm} \hat{R}_{mj} \nonumber \\
                       &\quad - \tfrac{2}{3} \hat{S}_{kl} \hat{S}_{lm} \hat{R}_{mo} \hat{R}_{ok} \delta_{ij}\ , \nonumber \\
    \hat{T}^{(10)}_{ij} &= \hat{R}_{ik} \hat{S}_{kl} \hat{S}_{lm} \hat{R}_{mo} \hat{R}_{oj} - \hat{R}_{ik} \hat{R}_{kl} \hat{S}_{lm} \hat{S}_{mo} \hat{R}_{oj}\ .
\end{align*}

Direct injection of the fully explicit closure \( \nabla \cdot a_{ij} \) into the RANS equations can lead to numerical stiffness and instability, particularly at high-Reynolds numbers or on under-resolved meshes. To mitigate this issue, ~\citet{mcconkey2024realizability} proposed a hybrid closure approach in which the linear term \( g_1 \hat{T}^{(1)}_{ij} \)---equivalent to the LEVM contribution---is treated implicitly by blending it with the eddy viscosity, whereas the functional forms of the remaining nonlinear terms \( g_2 \) through \( g_{10} \) are modeled using a machine-learning scheme and the resulting formulation of the nonlinear part of $a_{ij}$ is injected explicitly (or, directly) as a source term into the RANS equations, so
\begin{equation}
a_{ij} = -2 \nu_t S_{ij} + 2k \sum_{n=2}^{10} g_n \hat{T}^{(n)}_{ij}\ .
\end{equation}

This hybrid decomposition strikes a balance between numerical stability and model expressivity. The linear part retains the stability benefits of LEVMs, while the nonlinear basis terms allow the model to correct anisotropy-induced flow features. In the present work, we adopt this hybrid injection framework (referred orignally as the modified TBNN framework~\cite{mcconkey2024realizability} and investigate its performance in conjunction with the TBKAN architecture across several benchmark anisotropic flows.

\subsection{Tensor basis decomposition and feature invariants}

The TBNN framework~\cite{Ling2016} represents the anisotropy tensor \( a_{ij} \) as a linear combination of ten invariant tensor basis functions \( \hat{T}^{(n)}_{ij} \), each weighted by a scalar coefficient \( g^{(n)} \). This expansion, based on the Cayley–Hamilton theorem, guarantees frame invariance and allows for a physically interpretable decomposition of the anisotropic stresses.


In the original TBNN formulation, the coefficients \( g^{(n)} \) are modeled as functions of five scalar invariants \( \lambda_1 \) through \( \lambda_5 \) extracted from the normalized mean strain-rate tensor \( S_{ij} \) and mean rotation-rate tensor \( R_{ij} \). While this representation ensures Galilean and rotational invariance, it limits the expressivity of the model to specific features strictly derived from the local velocity gradient tensor.

In contrast, the realizability-informed TBNN framework proposed by \citet{mcconkey2024realizability} expands the input space to include additional physically motivated input features. This approach is adopted in the present study to provide the TBKAN model with a more diverse input feature set, which has been shown to improve generalization across anisotropic flow regimes.

Specifically, the input feature vector used in this work consists of two categories of features. First, we include heuristic features derived from the baseline $k$–$\omega$ SST RANS simulation. These include wall-distance-based Reynolds numbers, turbulence intensity ratios, and blending function indicators (e.g., \( q_1 \) through \( q_6 \)) that are commonly used within eddy-viscosity model formulations. These features encapsulate the localized flow behavior, such as near-wall effects, transition zones, and turbulence production.

Second, we incorporate a set of scalar invariants systematically constructed from a minimal integrity basis formed using \( S_{ij} \) and \( R_{ij} \). A symbolic computation (computer algebra) was used to eliminate any invariants that evaluate exactly to zero as a consequence of symmetry and other physical contraints in the (mean) flow (e.g., two dimensionality, the streamwise invariance of the flows etc.). This ensures that the final set of invariant features contributes meaningful variation across the dataset and remains physically relevant for the flows studied.

All input features are Galilean and rotationally invariant and are standardized using the mean and standard deviation from the training dataset prior to their use in neural network model training. This input feature set mirrors that used in~\citet{mcconkey2024realizability}. The combination of scalar invariants obtained here allows the TBKAN to leverage both physical modeling insights and data-driven input feature selection, thereby enhancing the flexibility and generalizability of the anisotropy mapping.

\subsection{Coefficient prediction using KAN}

The KAN replaces fixed nonlinear activation functions applied at the nodes with learnable univariate B-spline functions applied to sparse linear projections of the input at the edges between nodes, as described in Section~\ref{sec:theory}. This B-spline-based structure introduces a different inductive bias compared to MLPs, offering a locally adaptive function approximation that may provide a more parsimonious functional form for the representation of the scalar coefficient functions \( g^{(n)} \) in the tensor basis expansion of the anisotropy tensor.

The adoption of KANs is motivated by their ability to flexibly capture localized functional relationships in the input space, which is particularly relevant in turbulence modeling where quantities can span several orders of magnitude. Moreover, because KANs are able to represent complex mappings with fewer layers and potentially fewer input interactions, we hypothesize that they may require a reduced number of input features to achieve comparable or improved performance relative to MLPs. This property is of particular interest in the context of turbulence closure modeling, where the input feature space can be high-dimensional and highly correlated.

\subsection{Loss function with realizability penalty}
The loss function $\mathcal{L}$ used in training the TBKAN model utilizes the same form as described in~\citet{mcconkey2024realizability}; namely,
\begin{equation}
\mathcal{L} = \left\| a_{ij} - a_{ij}^\theta \right\|^2 + \alpha \cdot \mathcal{R}(b_{ij})\ ,
\label{eloss}
\end{equation}
where \( a_{ij} \) is the predicted anisotropy tensor (obtained from the data-driven modeling), \( a_{ij}^\theta \) is the target anisotropy tensor (obtained from a high-fidelity data such from DNS), and \( \mathcal{R}(b_{ij}) \) is a realizability penalty applied to the reconstructed Reynolds stress tensor. The realizability function penalizes eigenvalues of the normalized Reynolds stress tensor \( b_{ij} \equiv a_{ij}/(2k) \) that fall outside the physically admissible domain defined by Lumley's triangle ~\cite{lumley1976realizability}. This ensures that the predicted Reynolds stress tensor is positive semi-definite and, hence, is physically realizable~\cite{mcconkey2024realizability}.

The weight (regularization parameter) \( \alpha \) in Eq.~(\ref{eloss} is set to a value of 100 in this study, following empirical tuning in prior work, which balances the predictive accuracy with the enforcement of the physical realizability of the anisotropy tensor. This penalty improves the numerical stability of downstream RANS simulations and has been shown to reduce the frequency of unphysical predictions that can otherwise arise in high-gradient regions of turbulent flow fields.

\subsection{\label{sec:datasets}Datasets and input features}

Two datasets are used in this study: namely, the square duct and periodic hills flow datasets. For completeness, we provide a brief description of the square duct and periodic hills datasets, which are summarized in Table~\ref{tbl:datasets_trimmed}.  Here, \( Re_H \) is the Reynolds number of the square duct flow based on the duct half-height \( H \) and the bulk velocity \( U_b \), so \( Re_H = U_b H/\nu \), where \( \nu \) is the kinematic viscosity. In the periodic hills case, \( \alpha \) refers to the steepness parameter used to define the hill geometry~\cite{Xiao2020}. A larger value of \( \alpha \) corresponds to a steeper hill shape, increasing the severity of the flow separation and reattachment.

\begin{table}[htb]
\caption{\label{tbl:datasets_trimmed}Datasets used for training, validation, and testing for square duct and periodic hills flow cases.}
\begin{ruledtabular}
\begin{tabular}{lcc}
 & Square duct & Periodic hills \\
\hline
Number of data & 147,456 & 73,755 \\
\hfill \\
Parameter & $Re_H$ & $\alpha$ \\
\hline
Training set & \begin{tabular}[c]{@{}c@{}}1100--1600, 2205--3500\end{tabular} & 0.5, 1.0, 1.5 \\
Validation set & 1300, 1800, 3200 & 0.8 \\
Test set & 2000 & 1.2 \\
\end{tabular}
\end{ruledtabular}
\end{table}

The input features used to train the TBKAN and modified TBNN models in this study are derived from the baseline $k$–$\omega$ SST RANS simulations and are designed to ensure rotational and Galilean invariance in the predictions of the anisotropy tensor. The heuristic input features used in the neural network training include a wall-distance-based Reynolds number, a turbulence timescale ratio defined as \( k / (\epsilon + \omega) \), and six blending functions \( q_1 \)--\( q_6 \) commonly used in the formulation of the $k$-$\omega$ SST turbulence closure model. Furthermore, an additional 47 scalar invariants, extracted from the full set of invariants proposed by Wu et al.~\cite{Wu2018}, were used---scalar invariants from this full set that vanish (identically zero) owing to either the two-dimensionality of the flow or symmetry present in the flow were excluded.

The final feature set consists of 94 input quantities in total for the different flows. These were selected to ensure consistency across different flow regimes and to capture physically relevant turbulence characteristics. A complete listing and explanation of all input features is provided in Appendix~\ref{appendix:features} for transparency and reproducibility.


\begin{figure*}
\includegraphics[width=0.8\linewidth]{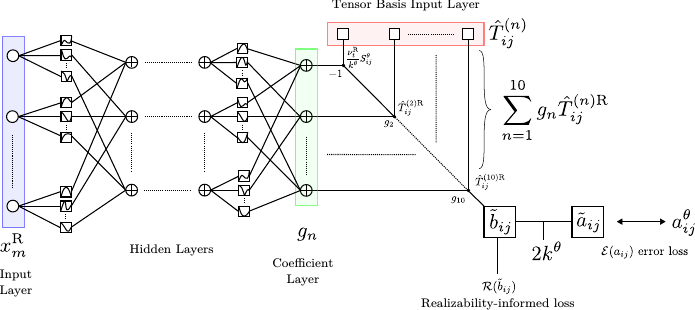}
\caption{\label{fig:tbkan-1}TBKAN architecture and training workflow. The training data consist of high-fidelity anisotropy tensor $a_{ij}^\theta$ and turbulence kinetic energy $k^\theta$ obtained from DNS.}
\end{figure*}

\begin{figure*}[htb]
\includegraphics[width=0.8\linewidth]{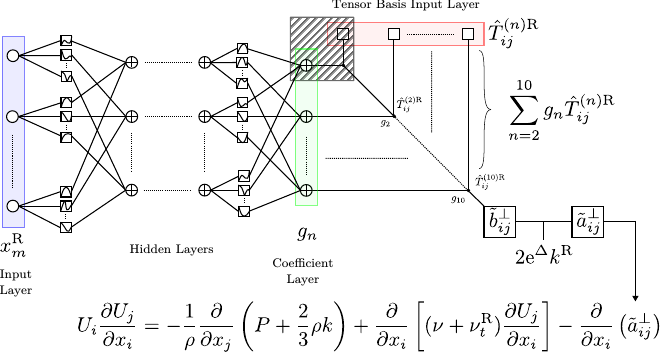}
\caption{\label{fig:tbkan-2}TBKAN injection into a fixed-anisotropy RANS solver using KCNN-predicted \( k \).}
\end{figure*}

\subsection{Training procedure and hyperparameter tuning}
\label{sec:training_procedure}

All models in this study, including the baseline modified TBNN and the proposed TBKAN architecture, were trained using the same case-wise groupings, training-validation-test splits (cf.~Table~\ref{tbl:datasets_trimmed}), and preprocessing pipeline as detailed in Section~\ref{sec:datasets}. The focus of this section is on the square duct and periodic hills datasets, which present strong anisotropy and serve as challenging benchmarks for Reynolds stress closure modeling.

Training was performed using the AMSGrad variant of the Adam optimizer~\cite{reddi2018convergence} with early stopping based on the validation loss. The batch size was set to 128, and all input features were standardized (normalized) using the training set mean and standard deviation.

The architecture of KAN is determined by a number of hyperparameters that need to be carefully chosen: namely, the network width, the spline grid size, and the order of the B-spline functions. To optimize these hyperparameters for a particular dataset, we employed Bayesian optimization (BO)~\cite{snoek2012practical}, a probabilistic global optimization technique that uses a surrogate model (typically Gaussian process regression) of the loss function to explore the hyperparameter space efficiently. BO is particularly suited for scenarios where training is computationally expensive, as it balances exploration and exploitation through the use of an acquisition function (e.g., expected improvement, probability of improvement, entropy search). To effectively apply Bayesian optimization, we first generated a set of initial evaluations using random search across the hyperparameter space. This warm-up strategy allowed the optimizer to gather informative samples from various regions of the design space, capturing a diverse range of model performance in the hyperparameter space across the different datasets and model configurations.

Table~\ref{tbl:hyperparameters_tbkan} summarizes the final training configurations selected using the hyperparameter tuning for both the modified TBNN and TBKAN models on the square duct and periodic hills flow datasets. For the modified TBNN model, the architecture selected consisted of a MLP with 7 hidden layers, each layer consisting of 30 neurons---giving a ($7\times30$ structure)---for both the square duct and periodic hills flow cases. For TBKAN, the "Configuration" in Table~\ref{tbl:hyperparameters_tbkan} refers to the number of nodes used for each hidden layer of the neural network (viz., the effective network width in each hidden layer). For example, the configuration $[8, 7, 9]$ used for the square duct flow case corresponds to a TBKAN model with 3 hidden layers consisting of 8, 7, and 9 nodes; similarly, for the periodic hills flow case, the optimal architecture found from the hyperparameter tuning has a configuration of $[8, 8, 8]$, implying a network with 3 hidden layers with each layer composed of 8 nodes. All these architectures were carefully selected based on their predictive performance exhibited in the hyperparameter tuning, specifically with respect to their minimization of the mean-squared-error (MSE) for the anistropy tensor.

\renewcommand{\arraystretch}{1.3}  
\begin{table}[htb]
\caption{\label{tbl:hyperparameters_tbkan}Optimal hyperparameters used in the training of the modified TBNN and TBKAN architectures for the square duct and periodic hills (PH) datasets.}
\begin{ruledtabular}
\begin{tabular}{lcccc}
Model & Dataset & \shortstack{Learning\\Rate} & Epochs & Architecture \\
\hline
TBNN  & Duct  & $5 \cdot 10^{-4}$ & 100   & MLP (7×30) \\
TBKAN & Duct  & $1 \cdot 10^{-4}$ & 375   & Configuration = [8, 7, 9] \\
\hline
TBNN  & PH    & $1 \cdot 10^{-5}$ & 19140 & MLP (7×30) \\
TBKAN & PH    & $1 \cdot 10^{-5}$ & 5153  & Configuration = [8, 8, 8] \\
\end{tabular}
\end{ruledtabular}
\end{table}

\subsection{Tensor basis Kolmogorov-Arnold network (TBKAN)}

As a recap, the tensor basis Kolmogorov-Arnold network builds directly upon the modified TBNN framework introduced in~\citet{mcconkey2024realizability}, in which the anisotropy tensor is predicted using an invariant-based tensor basis expansion. In the original formulation, a MLP was used to predict the scalar coefficients \( g^{(n)} \) corresponding to each tensor basis component \( \hat{T}^{(n)}_{ij} \). In the TBKAN model, this MLP is replaced by a KAN~\cite{liu2024kan}, which serves as a more structured and interpretable function approximator.


In the TBKAN architecture, the structure of the tensor basis model remains unchanged: the anisotropy tensor \( a_{ij} \) is expressed as a linear combination of invariant tensor basis functions. However, the KAN is now used for the modeling of the scalar coefficients \( g^{(n)} \) using a set of scalar invariants described previously in Section \ref{sec:training_procedure}. This replacement of MLP with KAN introduces a more compact and structured inductive bias, which improves smoothness, reduces model complexity, and retains the rotational and Galilean-invariant structure of the original TBNN proposed by \citet{Ling2016}.

The TBKAN training procedure closely mirrors that for the modified TBNN~\cite{mcconkey2024realizability}. The network is trained to minimize a loss function that includes a realizability penalty term on the reconstructed Reynolds stress tensor \( \tau_{ij} = 2k a_{ij} + \frac{2}{3}k \delta_{ij} \). The scalar invariants used as input features are computed from a baseline RANS simulation (using the $k$-$\omega$ SST turbulence closure model in our case).  The data used for the training consists of high-fidelity DNS data for the anisotropy tensor \( a_{ij} \). Figure~\ref{fig:tbkan-1} illustrates the overall architecture and the process used for training the TBKAN.

During inference, the trained TBKAN model is used to predict the anisotropy tensor \( \tilde{a}_{ij} \) from a new RANS solution, typically for an unseen flow condition or geometry. The predicted anisotropy tensor is then injected into a fixed-anisotropy RANS solver, replacing the Boussinesq approximation in the turbulence closure. This is done by holding \( \tilde{a}_{ij} \) fixed while iterating the mean momentum equations to convergence. The turbulent kinetic energy \( k \) used in the reconstruction of the Reynolds stress tensor is taken from a turbulence kinetic energy corrected neural network (KCNN) model trained separately, as described in~\citet{mcconkey2024realizability}. The resulting mean velocity fields and secondary flow patterns are then compared against available DNS test data. The injection procedure is depicted in Fig.~\ref{fig:tbkan-2}.


The injection framework used in this study builds directly on the modified TBNN approach proposed by~\citet{mcconkey2024realizability}, which incorporates both realizability constraints and a hybrid injection strategy to stabilize RANS predictions. In this setup, the Reynolds stress anisotropy tensor \( \tilde{a}_{ij} \) is predicted using either the modified TBNN or TBKAN model and injected into the RANS solver. To further improve the numerical stabiity of the injection, the linear eddy-viscosity component (corresponding to the \( g_1 \hat{T}^{(1)}_{ij} \) basis term) is retained and blended with the baseline model, while the nonlinear components of the modeled anisotropy tensor are injected explicitly. Additionally, the turbulent kinetic energy \( k \) is predicted using a separate $k$-correcting neural network (KCNN), which accounts for possible discrepancies in \( k \) between RANS and DNS. The prediction of \( k \) based on KCNN is used to reconstruct the full Reynolds stress tensor \( \tau_{ij} = 2k a_{ij} + \frac{2}{3}k\delta_{ij} \). In summary, the modified TBNN (or, TBKAN) model for the anisotropy tensor predictions and the KCNN model for the predictions of the turbulent kinetic energy correction constitute the complete machine learning-based closure framework used in this study.

The KCNN model is a compact neural network trained to correct the turbulent kinetic energy \( k \) field obtained from baseline RANS simulations. As detailed in~\citet{mcconkey2024realizability}, discrepancies in \( k \) between RANS and DNS were found to degrade the accuracy of reconstructed Reynolds stresses, even when the anisotropy tensor \( a_{ij} \) was well predicted. To mitigate this, the KCNN learns a mapping from local flow features (e.g., gradients of \( k \), \( \omega \), and wall distance metrics) to DNS-referenced \( k \) values. This corrected \( k \) field ensures consistent reconstruction of \( \tau_{ij} \), particularly in near-wall and recirculation regions where RANS often underpredicts turbulent energy. In all test cases, the KCNN is trained independently from the TBKAN and used only during the {\it a posteriori\/} injection step. The implementation of the TBKAN/KCNN framework used in this study is available on GitHub
\cite{kalia2025kanTBNN}.


\section{\label{sec:results}Results}

This section presents the evaluation of the TBKAN architecture on two benchmark turbulent fluid flow cases: namely, square duct and periodic hills flows. These cases are characterized by strong turbulence anisotropy, secondary flow formation, and separation phenomena, making them challenging test problems for traditional RANS closure models. We also note that the performance of TBKAN on the canonical flat plate boundary layer was previously investigated~\cite{mcconkey2024realizability}. In that study, it was demonstrated that TBKAN achieved comparable accuracy to a MLP-based modified TBNN using significantly fewer parameters, while also producing Reynolds stress predictions with improved realizability. Building on those findings, the present work extends the TBKAN approach to more complex flow configurations involving highly anisotropic and non-equilibrium flow regimes.

\subsection{\label{sec:square_duct}Square duct flow}

Turbulent flow through a square duct presents a well-established benchmark for evaluating the performance of turbulence models under strong anisotropy. The primary challenge lies in the accurate prediction of secondary flows that arise purely from turbulence-driven stress imbalances---a complexity in the flow phenomena that linear eddy-viscosity models (e.g., $k$–$\omega$ SST turbulence closure model) inherently fail to capture. The square duct test case used here follows the same setup as in previous work \cite{mcconkey2024realizability} which uses the DNS dataset provided by Pinelli et al.\cite{Pinelli2010} and the baseline RANS solution obtained by~\citet{McConkeySciDataPaper2021}.

\begin{figure}[htbp]
\centering
\includegraphics[width=0.95\linewidth]{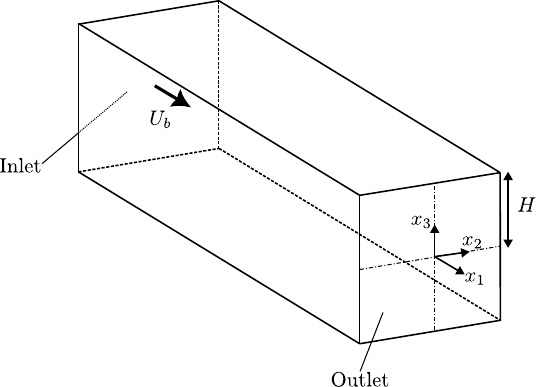}
\caption{Computational domain for the square-duct test case adapted from \citet{mcconkey2024realizability}.}
\label{fig:duct_domain}
\end{figure}

\begin{figure}[htbp]
\centering
\includegraphics[width=0.6\linewidth]{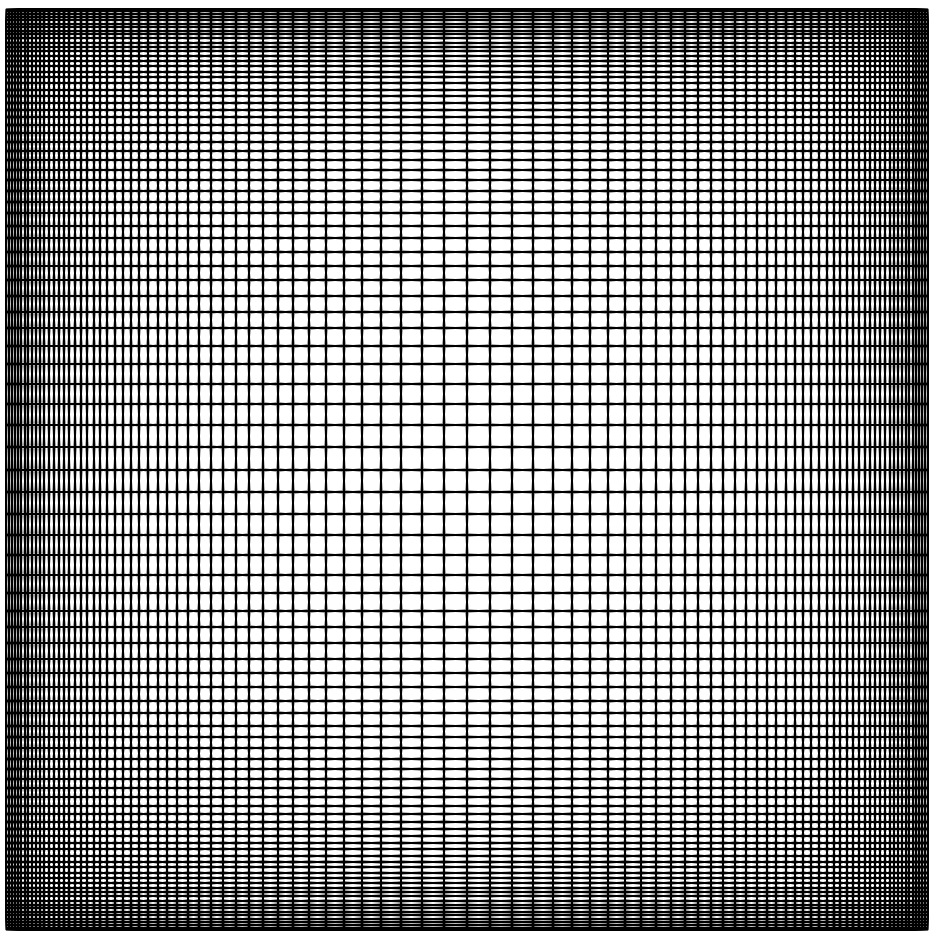}
\caption{The RANS mesh used for the square-duct simulation adapted from \citet{mcconkey2024realizability}.}
\label{fig:duct_mesh}
\end{figure}

The Reynolds number for the square duct flow is defined based on the duct half-height \( H \) and bulk velocity \( U_b \), so $Re_H \equiv U_b H/{\nu}$ where $\nu = 5 \times 10^{-6}$~m$^2$ s$^{-1}$ is the kinematic viscosity of the fluid. The computational domain used for the square-duct flow simulation is depicted in Fig.~\ref{fig:duct_domain}. The bulk velocity is varied in order to change the Reynolds number---the non-dimensional wall-normal distance $y^+ \equiv y u_\tau/\nu$ ($u_\tau$ is the friction velocity) of the first grid node from the wall verified \( y^+ \leq 1 \) for the (fixed) geometry and grid resolution in all the simulations, implying that the grid is fine enough to capture the near-wall region accurately. More specifically, the computational mesh used for the square-duct flow simulations is shown in Fig.~\ref{fig:duct_mesh}.

The TBKAN model was trained to predict the anisotropy tensor \( a_{ij} \) using scalar invariants described earlier (see Appendix~\ref{appendix:features}) within the framework of the tensor basis expansion developed by~\citet{Pope1975} as described in Sec.~\ref{sec:methodology}. Given the size and complexity of the square-duct dataset, a two-phase hyperparameter tuning strategy was utilized in order to ensure stable convergence and robust generalization of the results.

The first phase consisted of the evaluation of 20 model configurations selected using a random search---this search involved the variation of the network width, spline grid size, and learning rate. These exploratory runs were used to determine the viable regions of the hyperparameter space with an acceptable convergence and validation performance of TBKAN. The best-performing configurations were then used to initialize a second phase consisting of a Bayesian optimization, which was conducted for an additional 50 trials. The BO procedure minimized the MSE averaged across the normal (diagonal) and shear (off-diagonal) components of the anisotropy tensor.

\begin{figure}[htbp]
\centering
\includegraphics[width=0.9\linewidth]{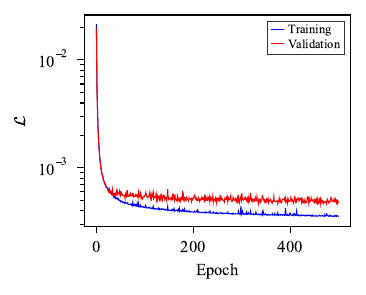}
\\[4pt] 
\textbf{(a)} TBKAN loss history
\\[10pt]
\includegraphics[width=0.9\linewidth]{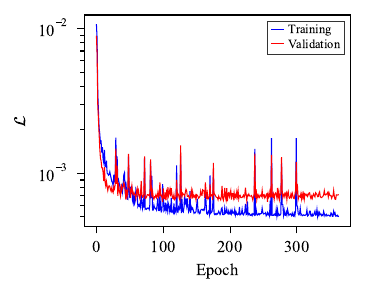}
\\[4pt]
\textbf{(b)} Modified TBNN loss history \cite{mcconkey2024realizability}
\caption{\label{fig:training-loss-comparison}Training and validation loss histories for the square-duct case. (a) TBKAN loss history exhibits smooth convergence. (b) Modified TBNN loss history achieves a lower final loss, albeit using a more complex model configuration.}
\end{figure}

The final selected TBKAN model consisted of a five-layer architecture with a width configuration of \([16, 8, 7, 9, 10]\), a spline grid size of 7 control points per edge, and a B-spline order of three (viz., cubic B-splines were chosen in the hyperparameter tuning). Note that the configuration specification here includes the number of nodes used in the input and output layers as well as the three hidden layers of the network. The model was trained using the AMSGrad optimizer with a learning rate of \( 1 \times 10^{-4} \), a mini-batch size of 64, and an early-stopping criterion based on the validation loss. This configuration demonstrated smooth training dynamics with minimal overfitting. While the baseline modified TBNN model achieved a lower MSE, the TBKAN architecture offered an enhanced interpretability and an improved convergence behavior. Figure~\ref{fig:training-loss-comparison} compares the training and validation loss histories for both the TBKAN and modified TBNN models on the square-duct case. The TBKAN model (cf.~Fig.~\ref{fig:training-loss-comparison}(a)) exhibits a smooth convergence with minimal indications of overfitting, while the modified TBNN baseline model (cf.~Fig.~\ref{fig:training-loss-comparison}(b)) is observed to achieve a slightly lower final validation loss, but at the cost of an increased model complexity. To avoid overfitting, an early-stopping criterion was applied during training.
To this purpose, the model’s performance is monitored on the validation set during training and convergence to the optimal solution is selected at that point in the training process when the average change in the validation metric is constant over a pre-selected number of epochs---early stopping trigger used to prevent overfitting. 
This strategy ensured a fair comparison across model architectures and helped mitigate issues arising from noisy validation data in high-dimensional parameter spaces. 

\begin{figure*}[htbp]
\includegraphics[width=\linewidth]{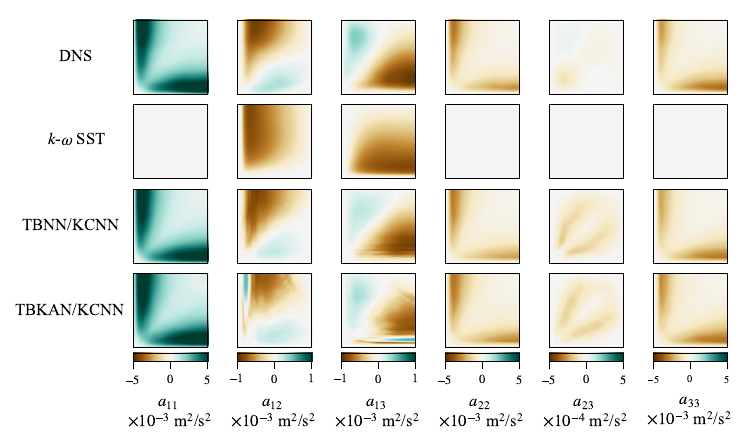}
\caption{\label{fig:sd-apred}Predicted components \( a_{ij} \) of the anisotropy tensor obtained from the TBKAN/KCNN model in comparison with the baseline RANS predictions using the $k$-$\omega$ SST model, the reference DNS data, and the predictions obtained with the modified TBNN/KCNN model. All anisotropy components are shown for the vertical cross-section in the square duct at \( x_1 = 0 \).}
\end{figure*}

To evaluate the {\it a priori\/} performance of the TBKAN model, Fig.~\ref{fig:sd-apred} presents the predicted anisotropy tensor components \( a_{ij} \) for the vertical cross-section at \( x_1 = 0 \). This figure compares the TBKAN/KCNN model predictions of $a_{ij}$ with those obtained from the reference DNS data as well as with the baseline RANS predictions obtained with the $k$-$\omega$ SST model and the predictions obtained from the modified TBNN/KCNN model. 
Recall that the corrections to turbulent kinetic energy \( k \) was predicted using a separate KCNN model. Since \( a_{ij} \equiv 2k b_{ij} \), an accurate estimation of \( k\) is essential for the proper reconstruction of the dimensional anisotropy tensor. We note that the KCNN recovers the near-wall \( k \) distribution typically underpredicted by the RANS $k$-$\omega$ SST model---indeed, this aspect of the data-driven modeling is important as it enables a stable and accurate injection of the anisotropy tensor into the  RANS model for {\it a posteriori\/} predictions of the velocity field (see below).

A careful perusal of Fig.~\ref{fig:sd-apred} shows that the TBKAN/KCNN model predictions clearly exhibit a significantly improved agreement with DNS in comparison to the baseline RANS predictions based on the $k$-$\omega$ SST model, particularly in the \( a_{11} \) component near the duct walls and corners, where anisotropy effects are most pronounced. The model also captures key structural features in off-diagonal components such as \( a_{12} \) and \( a_{23} \), albeit with a slight underprediction in regions of low magnitude in comparison with the reference DNS results. Taken together, these observations underscore the ability of the TBKAN/KCNN model to learn spatially accurate representations of the turbulence anisotropy in a square duct.

\begin{figure}[htbp]
\centering
\includegraphics[width=\linewidth]{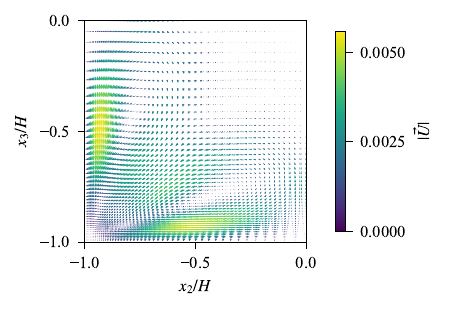}
\caption{\label{fig:sd-vectors-pred-kan}Predicted secondary velocity vectors \( (U_2, U_3) \) of the fully-developed flow in a vertical cross-section through the square duct at \( x_1 = 0 \)  obtained from the TBKAN/KCNN injection of the anisotropy tensor into a RANS model.}
\end{figure}

The predictions of the anisotropy tensor $a_{ij}$ obtained with TBKAN/KCNN were injected into a RANS model. As shown in Fig.~\ref{fig:sd-vectors-pred-kan}, the RANS model with this injection of $a_{ij}$ correctly predicted the formation of the mean secondary flow field in a vertical cross-section of the square duct (viz., in a plane perpendicular to the primary flow direction in the duct). We note that Fig.~\ref{fig:sd-vectors-pred-kan} only exhibits the flow field in the lower-left quadrant of the square duct. The flow pattern displayed in this {\it quadrant\/} of the square duct consists of a corner vortex pair that exhibits the characteristic two-fold symmetry about the corner bisector---an important characteristic feature of the secondary flow (of the secondary kind) in a square duct that cannot be predicted using the baseline RANS model with a $k$-$\omega$ SST turbulence closure.

\begin{figure*}[htbp]
\centering
\subfloat[$U_2$ velocity profile\label{fig:duct_profiles_U2}]{
    \includegraphics[width=0.475\linewidth]{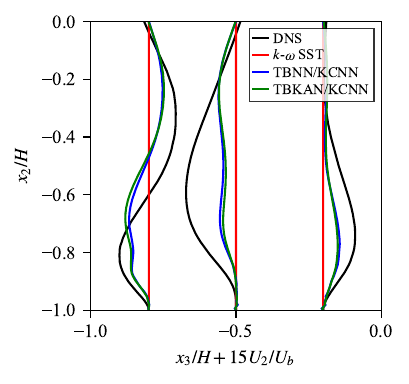}
}
\hfill
\subfloat[$U_3$ velocity profile\label{fig:duct_profiles_U3}]{
    \includegraphics[width=0.475\linewidth]{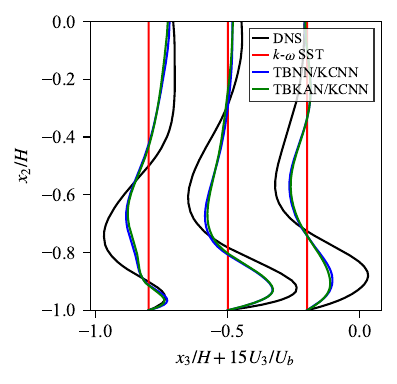}
}
\caption{\label{fig:duct_velocity_profiles}Velocity profiles of the in-plane components \( U_2 \) and \( U_3 \) in a plane perpendicular to the primary (streamwise) flow direction in a square duct. RANS model predictions obtained using the TBKAN/KCNN, modified TBNN/KCNN, and baseline $k$-$\omega$ SST turbulence closure models are compared with the reference data for the mean velocity obtained using DNS.}
\end{figure*}

Figure~\ref{fig:duct_velocity_profiles} displays the plots of vertical profiles of the in-plane velocity components \( U_2 \) and \( U_3 \) at various spanwise locations in a plane perpendicular to the streamwise flow direction, comparing RANS model predictions obtained using the TBKAN/KCNN, modified TBNN/KCNN, and baseline $k$-$\omega$ SST turbulence closure models with the reference mean velocity data obtained from DNS. A careful examination of the figure shows that both the TBKAN/KCNN and modified TBNN/KCNN closure models exhibit improved agreement with DNS data for both components of the in-plane velocity in comparison with the baseline closure model, especially in the vortex-dominated corner regions, capturing the curvature and peak locations more accurately than the baseline model which (as indicated above) failed to predict the secondary mean velocity field in the square duct. However, it is noted that the ``strength'' of the weak secondary flow is underpredicted by both the TBKAN/KCNN and TBNN/KCNN closure models.

\begin{figure*}[htbp]
\includegraphics[width=\linewidth]{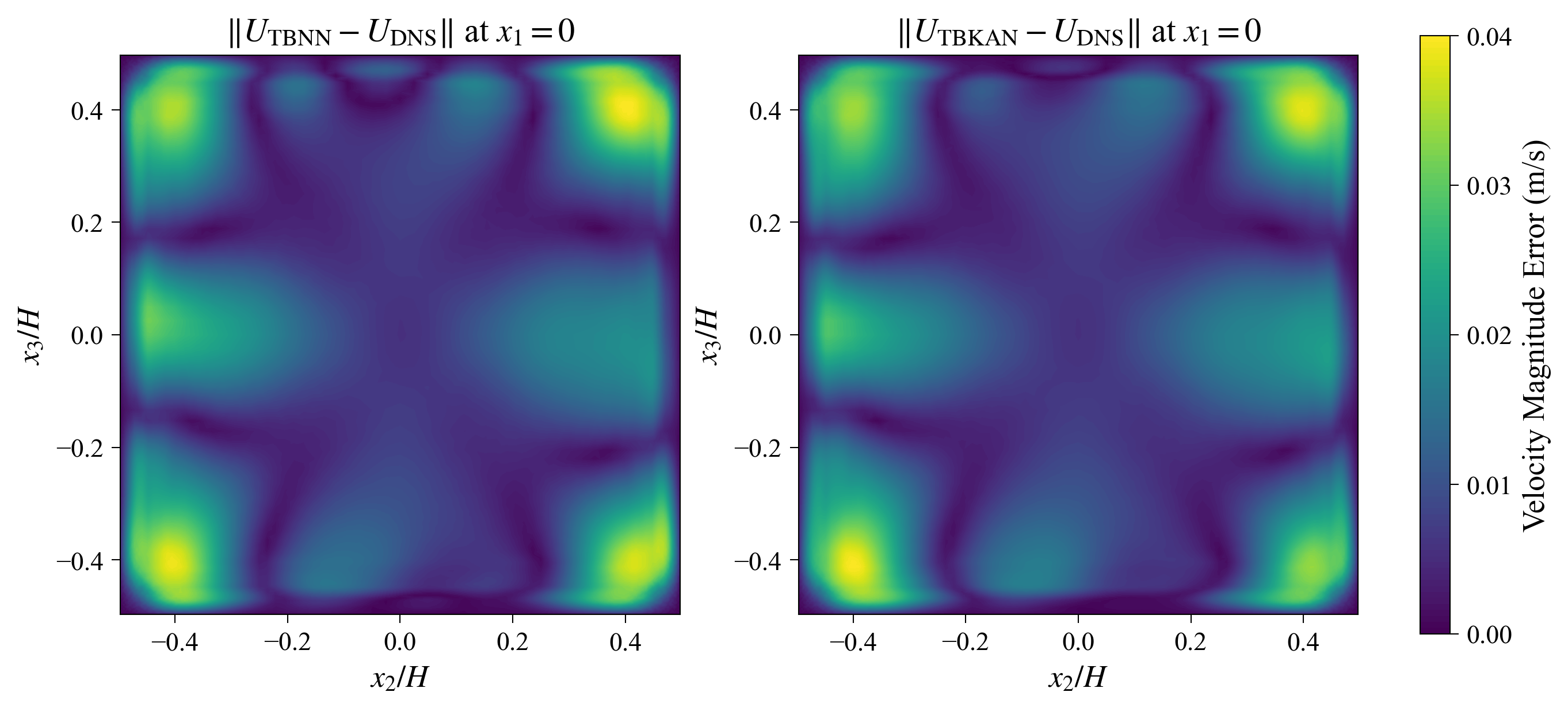}
\caption{\label{fig:errors-sd} Contours of the velocity magnitude error in a vertical plane perpendicular to the primary mean flow direction in a square duct at $x_1=0$: L2-norm of the vector difference between the TBNN/KCNN (left frame) and TBKAN/KCNN (right frame) predictions and the DNS reference data.}
\end{figure*}

Figure~\ref{fig:errors-sd} displays contour plots of the velocity magnitude error defined as the Euclidean norm of the vector difference between the velocity magnitude of the two in-plane velocity components obtained from RANS predictions using the modified TBNN/KCNN (left frame) and the TBKAN/KCNN (right frame) turbulence closure models with that obtained from the reference DNS velocity data of the same quantity. An examination of this figure shows that the velocity magnitude errors in the vertical plane perpendicular to the primary flow direction in the square duct are smallest at the core and along the two diagonals of the square duct. The largest velocity magnitude error occurs at the four corners of the square duct where it has already been shown previously that the two data-driven turbulence closure models underpredict the strength of the weak secondary flow, a defect that probably arises from an incorrect prediction of the secondary shear stress component that arises as a result of the generation of the secondary mean flow in a vertical cross-section of the square duct.

\subsection{\label{sec:periodic_hills}Periodic hills flow}

The flow over periodic hills is a widely studied benchmark for turbulence modeling due to the challenging physics it presents, including separation, adverse and favorable pressure gradients, and reattachment. The periodic hills configuration used in this study concerns a flow at a Reynolds number $Re_H \equiv U_b H/\nu = 5600$ ($U_b$ is the bulk velocity and $H$ is the hill height)---this case was investigated by~\citet{Xiao2020} using DNS. This dataset~\cite{Xiao2020}, as well as the RANS predictions for the same case provided by~\citet{McConkeySciDataPaper2021}, was used for training and evaluation of the neural networks used in modeling of the anisotropy tensor.

\begin{figure*}[htbp]
\centering
\includegraphics[width=\linewidth]{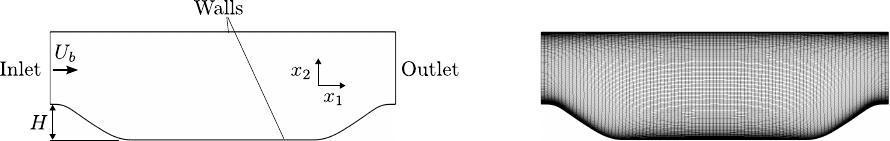}
\caption{\label{fig:periodic_hills_setup}Computational domain (left panel) and mesh (right panel) for the periodic hills flow case with a steepness \( \alpha = 1.2 \), adapted from ~\citet{mcconkey2024realizability}.}
\end{figure*}

The boundary conditions imposed on the computational domain of the periodic hills flow are as follows: a periodic condition is applied in the streamwise direction and a no-slip boundary condition is applied at the top and bottom walls of the computational domain. The geometry and mesh for the periodic hills flow case are depicted in Fig.~\ref{fig:periodic_hills_setup}. The hill shape is parameterized by a steepness factor \( \alpha \), where \( \alpha = 1.2 \) is used as the test configuration in this study.

After hyperparameter tuning, the best-performing TBKAN model for prediction of the anisotropy tensor for the periodic hills flow had a network configuration with widths of [17, 8, 8, 8, 10] and used a spline grid size of 5. Again, the configuration here includes the number of nodes used for the input and output layers as well as the three hidden layers. To identify an optimal TBKAN architecture, a comprehensive random search was conducted over a defined range of candidate configurations. Specifically, the number of nodes per hidden layer and the spline grid size were each varied between 5 and 15. This range was selected to mitigate the risk of overfitting—observed to be more prevalent at higher values—and to reduce the computational cost. The B-spline order was fixed at three (viz., a cubic B-spline), consistent with prior work on the application of TBKAN to a flat plate boundary layer flow~\cite{mcconkey2024realizability}. The learning rate was varied in the range \([1 \times 10^{-5},\ 5 \times 10^{-5}] \), with values in this interval found to provide stable convergence and robust generalization in the exploratory runs.

\begin{figure}[htbp]
\centering
\includegraphics[width=0.85\linewidth]{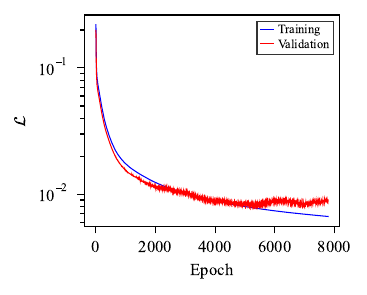}
\\[4pt]
\textbf{(a)} TBKAN loss history
\\[10pt]
\includegraphics[width=0.85\linewidth]{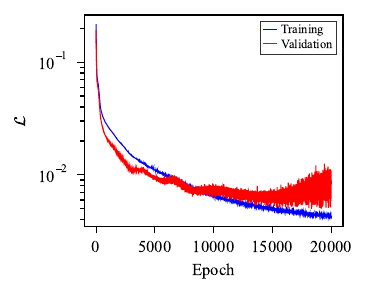}
\\[4pt]
\textbf{(b)} TBNN loss history \cite{mcconkey2024realizability}
\caption{\label{fig:losses-phll}Training and validation loss histories for the periodic hills flow case. (a) TBKAN loss history and (b) modified TBNN loss history.}
\end{figure}

Figure~\ref{fig:losses-phll} exhibits the training and validation loss histories for the periodic hills flow case for the best-performing TBKAN and modified TBNN models obtained from the extensive hyperparameter tuning conducted for each model. The loss function associated with the training of the TBKAN architecture (cf.~Fig.~\ref{fig:losses-phll}(a)) exhibits a smooth convergence towards the minimum loss solution over fewer epochs than that for the modified TBNN architecture, stabilizing around a consistent minimum without significant overfitting. In contrast, the modified TBNN model exhibits a more noisy behavior in the validation loss (cf.~Fig.~\ref{fig:losses-phll}(b)), despite reaching a slightly lower minimum loss compared to that of the TBKAN model---this is likely due to the larger number of parameters in the model which, in turn results in longer training times. To prevent overfitting and ensure a robust convergence to an optimal solution, both models were trained using an early-stopping criterion based on behavior of the validation loss--a patience threshold of 5000 epochs was applied to stop training if no improvement was observed in the validation loss.
This strategy allowed both the modified TBNN and TBKAN models to converge efficiently to optimal solutions while avoiding excessive training that could lead to instability and/or an overfitting of the training data which, in turn, can result in a degraded generalization performance of the models.

\begin{figure*}[htbp]
\centering
\includegraphics[width=\linewidth]{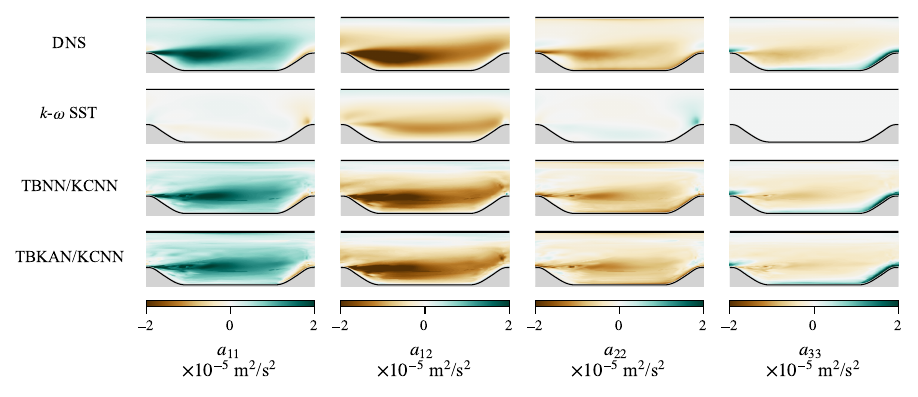}
\caption{\label{fig:train-phll}Streamwise development of four components of the anisotropy tensor (\( a_{11}, a_{12}, a_{22}, a_{33} \)) for the periodic hills flow case with a steepness parameter of \( \alpha=1.2 \). The predictions of these components obtained from the baseline $k$-$\omega$ SST, modified TBNN/KCNN, and TBKAN/KCNN models are compared with those obtained from a reference DNS dataset~\cite{Xiao2020}.}
\end{figure*}

Figure~\ref{fig:train-phll} presents an {\it a priori\/} comparison of the streamwise development of various components of the anisotropy tensor (viz., \( a_{11}, a_{12}, a_{22}, a_{33} \)) for the periodic hills test case with a steepness parameter of \( \alpha = 1.2 \) obtained from the baseline $k$-$\omega$ SST, modified TBNN/KCNN, and TBKAN/KCNN turbulence closure models. These predictions are compared to the reference DNS anisotropy tensor. The baseline $k$-$\omega$ SST (eddy-viscosity) model significantly underpredicts all components of the anisotropy tensor, particularly the normal stresses \( a_{11} \) and \( a_{33} \)---in particular, this model fails to capture both the magnitude and location of maximum of the anisotropy. In contrast, both machine-learning-based turbulence models are seen to provide significantly improved predictions of the anisotropy tensor components, with both the modified TBNN/KCNN and TBKAN/KCNN models displaying sharper transitions and markedly better agreement with the DNS results than the baseline $k$-$\omega$ SST model, especially in the separated and reattachment regions of the flow. Overall, the predictions provided by the TBKAN/KCNN model of the anisotropy tensor are in very good conformance with those provided by the modified TBNN/KCNN model, albeit the spline-based neural network has a simpler structure and uses fewer parameters than the MLP-based neural network. Moreover, a careful examination of Fig.~\ref{fig:train-phll} suggests that predictions of \( a_{12} \) and \( a_{33} \) obtained from the TBKAN/KCNN model are slightly more consistent with the DNS data within separation and reattachment zones of the flow than those provided by the modified TBNN/KCNN model, suggesting a slightly better representation of the shear-layer and normal-stress behavior in these regions. Despite the simplicity of the TBKAN/KCNN model, it is seen that this model provides high fidelity in its prediction of the turbulence characteristics in the periodic hills flow, with the potential of offering improved regularization and interpretability of the results.

\begin{figure*}[htbp]
\centering
\includegraphics[width=0.9\linewidth]{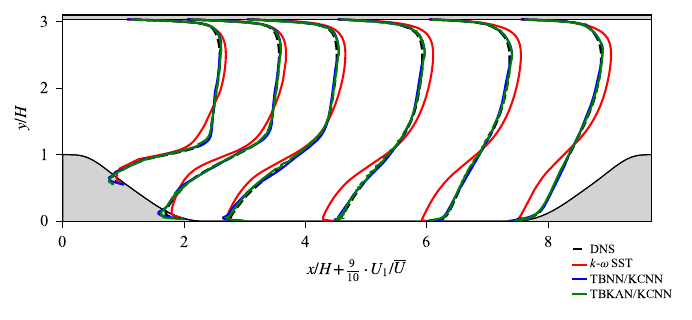}
\caption{\label{fig:phll-u1-multiline} Vertical profiles of the normalized streamwise velocity (\( U_1/\overline{U} \)) obtained at various streamwise locations (\( x/H \)) in the periodic hills flow ($\alpha=1.2$). The predictions of the normalized streamwise velocity provided by RANS simulations conducted with the baseline $k$-$\omega$ SST, TBNN/KCNN, and TBKAN/KCNN turbulence closure models are compared with those provided by a reference DNS dataset~\cite{Xiao2020}.
}
\end{figure*}

To evaluate the utility of the trained TBKAN/KCNN model beyond {\it a priori\/} predictions of the anisotropy tensor, we conducted an {\it a posteriori\/} analysis by injecting the predicted anisotropy fields provided by this model into the RANS model (mean momentum transport equation). 
This {\it a posteriori\/} assessment allowed an examination of whether the improved anisotropy predictions provided by the TBKAN/KCNN model lead to physically meaningful improvements in the prediction of the mean velocity field for the periodic hills flow case. Towards this objective, Fig.~\ref{fig:phll-u1-multiline} presents a detailed comparison of the vertical profiles of the normalized streamwise velocity \( U_1/\overline{U} \) at multiple streamwise locations in the periodic hills flow field for various RANS model predictions. More specifically, this figure compares the predictions of this quantity obtained from RANS simulations conducted with the baseline $k$–$\omega$ SST, modified TBNN/KCNN, and TBKAN/KCNN turbulence closure models to reference DNS data for the streamwise mean velocity.

A careful inspection of Fig.~\ref{fig:phll-u1-multiline} shows that the RANS model prediction using the modified TBNN/KCNN and TBKAN/KCNN turbulence closures exhibited significantly improved agreement with DNS data for all sampling stations and for all regions of the flow (e.g., separated shear layer and near reattachment zones) in comparison with the RANS model predictions obtained with the baseline $k$-$\omega$ SST (eddy viscosity) turbulence closure. Moreover, it is seen that the RANS predictions obtained with the $k$-$\omega$ SST model consistently underpredicts the velocity deficit and overestimates the flow recovery, while the RANS predictions obtained with both the modified TBNN/KCNN and TBKAN/KCNN models partially correct these trends---albeit these predictions are seen to exhibit greater deviations from the reference DNS data near the hill crest and in the recirculation regions of the flow. 

\begin{figure*}[htbp]
\centering
\includegraphics[width=\linewidth]{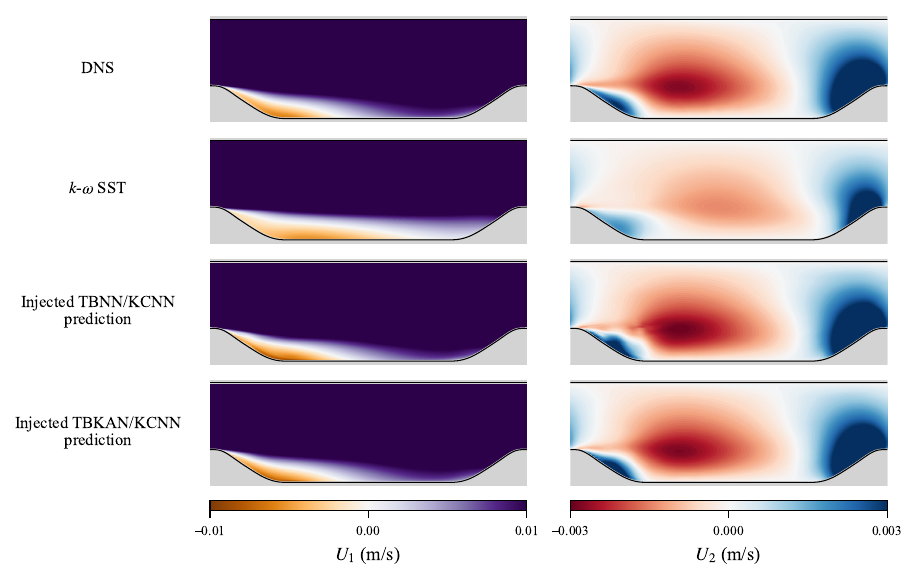}
\caption{\label{fig:u_inj_phll} Predictions of the streamwise $U_1$ (left panel) and vertical $U_2$ (right panel) mean velocity fields for the periodic hills flow test case ($\alpha=1.2$) obtained from RANS simulations conducted with the baseline $k$-$\omega$ SST, modified TBNN/KCNN, and TBKAN/KCNN turbulence closure models. The predictions are compared to the reference DNS data~\cite{Xiao2020} for the streamwise and vertical mean velocity fields.
}
\end{figure*}

Figure~\ref{fig:u_inj_phll} displays predictions of the mean streamwise \( U_1 \) and vertical $U_2$ velocity fields for a periodic hills flow case with a steepness parameter of $\alpha = 1.2$. The predictions for these two mean flow quantities were obtained from RANS simulations conducted with the baseline $k$-$\omega$ SST, modified TBNN/KCNN, and TBKAN/KCNN turbulence closure models and compared with the reference DNS data~\cite{Xiao2020}.  We note that the RANS predictions for both the streamwise and vertical mean velocity obtained with the two data-driven turbulence closure models are in significantly better conformance with the reference DNS data than those obtained with the baseline $k$-$\omega$ SST turbulence closure model. In particular, the predictions for the streamwise and vertical mean velocity obtained with the two former models provide better predictions of features of the flow in the recirculation zone. Furthermore, the location of the reattachment point in the flow obtained with the two former models are with better agreement with the DNS data than that obtained with the latter model. Finally, predictions provided by the two former models of the acceleration and deceleration zones induced by the geometry of the periodic hills are in better conformance with DNS results than those provided by the latter model, particularly along the separated shear layer.

\begin{figure*}[htbp]
\centering
\includegraphics[width=\linewidth]{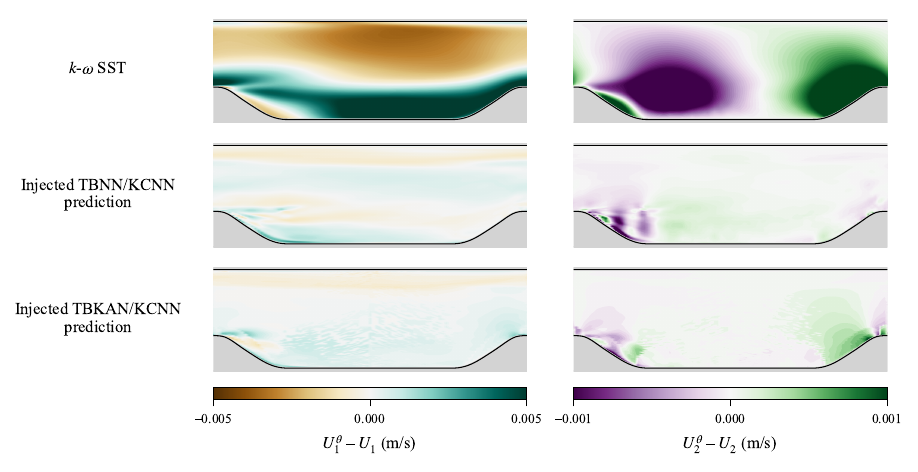}
\caption{\label{fig:phll_errors_U}Magnitude of streamwise $U_1$ (left frame) and vertical $U_2$ (right frame) velocity errors relative to the reference DNS data~\cite{Xiao2020} ($U_1^\theta$ and $U_2^\theta$) for the periodic hills flow case ($\alpha=1.2$) obtained from RANS predictions with the baseline $k$-$\omega$ SST, modified TBNN/KCNN, and TBKAN/KCNN turbulence closure models.}
\end{figure*}

To further quantify the model performance, contours of the error in the RANS predictions of the streamwise and vertical mean velocities relative to the reference DNS data~\cite{Xiao2020} were determined for the baseline $k$-$\omega$ SST, modified TBNN/KCNN, and TBKAN/KCNN turbulence closure models. The results of this analysis are presented in Fig.~\ref{fig:phll_errors_U}. It is seen that the errors obtained with the two data-driven turbulence closure models exhibited consistently small velocity errors throughout the complex flow regions associcated with the recirculation and reattachment zones. 
Furthermore, while some discrepancies in predictions of the mean flow persist near the hill crest (especially for the vertical mean velocity), the overall magnitude of the velocity errors is significantly lower for the two data-driven turbulence closure models than those for the baseline $k$-$\omega$ SST model.


\section{\label{sec:discussion}Discussions}

\subsection{\label{sec:model_comparison}Summary of model performance and complexity}

An important consideration when evaluating neural network architectures for turbulence modeling is the number of trainable parameters required to achieve a given level of predictive performance. For the square duct flow case, the modified TBNN architecture employed a MLP with 7 hidden layers of 30 neurons each. The input layer to the network had 16 neurons to accommodate the 16 input features and the output layer had 10 neurons to provide a prediction of the 10 tensor basis coefficients that define the anisotropy tensor. This results in 6,180 trainable weights computed as follows: the input layer contributes \( 16 \times 30 = 480 \) weights, each of the 7 hidden layers contributes \( 30 \times 30 = 900 \) weights (totaling \( 6 \times 900 = 5,400 \) weights contributed by the hidden layers), and the final output layer contributes \( 30 \times 10 = 300 \) weights. In addition, the network also uses 220 biases to give a total number of 6,400 trainable parameters for the modified TBNN architecture used to model the anisotropy tensor for the square duct flow.  Similarly, for the periodic hills flow, the modified TBNN architecture employed a MLP with the same hidden layer and output layer structure as that used for the square duct flow, but the input layer in this case consists of 17 neurons accommodating the 17 input features using in the training. In consequence, the total number of trainable parameters (weights and biases) for this neural network is 6,430. These neural network architectures are summarized in Table~\ref{tbl:hyperparameters_tbkan}. This explicit determination of the number of trainable parameters provides a reference point for the assessment of the model complexity of the modified TBNN architecture (based on MLP) in comparison with the TBKAN architecture (based on KAN).

Towards this objective, the TBKAN architecture is a B-spline-based function approximator and, as such, provides a more parsimonious representation of the mapping from the input features to the anisotropy tensor components in the sense that this mapping requires significantly fewer parameters for the characterization of the mapping. The TBKAN architecture used to model the anisotropy tensor for the square duct flow had an input layer with 16 nodes, three hidden layers with network widths of [8, 7, 9], and an output layer with 10 nodes and with a spline grid size of 7, this architecture involves 3,024 trainable parameters. Similarly, for the periodic hills flow, the TBKAN architecture used to model the anisotropy tensor had an input layer with 17 nodes, three hidden layers with widths of [8, 8, 8], and an output layer with 10 nodes, which with a spline grid size of 5, the architecture used 1,117 trainable parameters to parameterized the mapping. These comparisons highlight the significant reduction in model complexity achieved by the TBKAN architecture, while providing a comparable predictive accuracy to that provided by the modified TBNN architecture. In particular, the TBKAN architecture used between about one-half (square duct flow) and one-sixth (periodic hills flow) the number of learnable parameters to represent the mapping of the input features to the anisotropy tensor compared to that of the modified TBNN architecture, while providing a commensurate predictive performance (accuracy).

\renewcommand{\arraystretch}{1.3}  
\begin{table*}[htb]
\caption{\label{tbl:summary_errors_params}Comparison of model performance and complexity across the square duct and periodic hills flow test cases. Reported values include mean-squared-error (MSE) in the prediction of the anisotropy tensor  (\( a_{ij} \)) and the mean velocity field (\( U_i \)). The model performance is assessed in terms of the total number of learnable parameters required to achieve this performance.}
\begin{ruledtabular}
\begin{tabular}{l c c c c}
\textbf{Test Case} & \textbf{Model} & \textbf{MSE in \( a_{ij} \)} & \textbf{MSE in \( U_i \)} & \textbf{\# Learnable Parameters} \\
\hline
\multirow{2}{*}{Square Duct} 
    & TBNN/KCNN     & \(2.01 \times 10^{-9}\)  & \(3.01 \times 10^{-6}\)  & 6400 \\
    & TBKAN/KCNN    & \(3.39 \times 10^{-8}\)  & \(2.84 \times 10^{-6}\)  & 3024 \\
\hline
\multirow{2}{*}{Periodic Hills} 
    & TBNN/KCNN     & \(2.17 \times 10^{-12}\) & \(2.23 \times 10^{-8}\)  & 6430 \\
    & TBKAN/KCNN    & \(3.18 \times 10^{-12}\) & \(3.71 \times 10^{-8}\)  & 1117 \\
\end{tabular}
\end{ruledtabular}
\end{table*}

\subsection{Model efficiency and interpretability}

Table~\ref{tbl:summary_errors_params} summarizes the model performance (MSE) of the modified TBNN/KCNN and TBKAN/KCNN models for {\it a priori\/} predictions of the anisotropy tensor $a_{ij}$ and for {\it a posteriori\/} predictions of the mean velocity field $U_i$ (after injections of the models into a RANS framework) for the square duct and periodic hills flows. A perusal of Table~\ref{tbl:summary_errors_params} indicates that the TBKAN/KCNN architecture consistently achieves comparable or slightly improved \mbox{performance} in predicting both the anisotropy tensor \( a_{ij} \) and the mean velocity field \( U_i \), while using significantly fewer learnable parameters than the modified TBNN/KCNN architecture. This reduction in model complexity, which ranges from approximately 50\% fewer parameters for the square duct flow to over 80\% fewer parameters for the periodic hills flow highlights the efficiency (parsimony) of TBKAN's spline-based representation relative to the more conventional MLP representation used in TBNN. Moreover, the TBKAN/KCNN model shows a small but notable improvement in  the \emph{a posteriori} performance for the periodic hills flow, particularly in regions of flow separation, as reflected in a reduced MSE in the prediction of \( U_i \).

Each edge (connections between nodes) of a KAN-based model is associated with a learnable univariate (activation) function that is implemented as B-splines of a given order (viz., piecewise polynomials joined at knot points and parameterized by a set of learnable control points). In consequence, this representation is extremely flexible and can be used to approximate a wide range of shapes. In consequence, a KAN-based model potentially combines computational efficiency through a reduced model complexity with a potentially improved interpretability---the design of a KAN inherently provides a greater interpretability than a MLP owing to the fact that a univariate function (B-spline) provides an explicit representation of the relationship between a single input (or an intermediate sum) and its contribution to the next layer.

Owing to the presence of a learnable activation function on each edge of the network, it is feasible potentially to approximate the learned behavior of the aggregate network with concise mathematical expressions. In this sense, the TBKAN can serve as an interpretable front end for symbolic regression---this opens up exciting possibilities for extending TBKAN beyond the conventional neural network application as in the more conventional MLP used in TBNN. Because each edge of the network in TBKAN learns an explicit spline-based function, it becomes possible to approximate these mappings using symbolic regression, potentially enabling an efficient algebraic approximation of the underlying model for the various components of the anisotropy tensor without being constrained to predetermined functional forms. 

This can provide compact, closed-form algebraic expressions for the mapping from the input features to the anisotropy tensor, offering potentially deeper insight into the underlying flow physics and contributing to more interpretable (transparent) and generalizable algebraic stress models for turbulence closure. The derived algebraic stress model can be directly integrated into the RANS framework, without requiring the injection of a complex neural network.
In this way, TBKAN potentially provides the initial impetus to balance model efficiency with intepretability (transparency), setting the stage for future work that bridges data-driven and analytical modeling approaches for turbulence closure.

\section{\label{sec:conclusion}Conclusion}

This study extends the TBKAN architecture, previously applied to a flat-plate boundary-layer flow, to more complex turbulent flows characterized by strong turbulence anisotropy, secondary flow motion, and flow separation. Specifically, we evaluate the efficacy of TBKAN for modeling the square duct and periodic hills flows, which present substantially greater modeling challenges than flow over a flat plate. This investigation serves as a benchmark study for assessment of the robustness of data-driven turbulence closure schemes based on a KAN architecture and contrasts this strategy with one that uses a more conventional MLP architecture.

Unlike previous work, this study is the first to demonstrate {\it a posteriori\/} accuracy resulting from the injection of a TBKAN-predicted anisotropy tensor into the RANS framework for prediction of the mean velocity field. This enables a direct assessment of how the TBKAN-based turbulence closure impacts the predictive accuracy of the mean flow velocity in comparison to a conventional baseline turbulence closure scheme (e.g., $k$-$\omega$ SST model). Using a novel strategy for the injection of the predicted anisotropy tensor into the RANS framework in a highly stable manner and a realizability-informed loss function, TBKAN provides physically consistent and excellent predictions of the anisotropy tensor, as well as stable and good predictions for the mean velocity fields, including various flow features such as accurate prediction of the reattachment location in the periodic hills flow and the prediction of secondary flow motions (e.g., corner vortices) in the square duct flow.

Interestingly, the KAN-based TBKAN model achieves these results while requiring significantly fewer trainable parameters than the standard MLP-based TBNN model, highlighting the advantages of the KAN-based model as a more parsimonious and interpretable alternative to the more conventional MLP-based model. Moreover, the KAN-based model exhibits smooth training dynamics, good generalizability, and a strong ability to learn flow-specific corrections in challenging flow regimes.

These findings underscore the viability of KAN-based architectures in turbulence closure modeling and motivate future work focused on more complex three-dimensional flows, the incorporation of uncertainty quantification into the KAN model formulation, and further investigations using physics-informed loss functions to improve generalizability and reliability across diverse classes of flows.

\section*{Data Availability Statement}
The data that support the findings of this study are openly available in "A curated dataset for data-driven turbulence modelling" at \url{https://www.kaggle.com/datasets/ryleymcconkey/ml-turbulence-dataset}.

\appendix
\section{INPUT FEATURES USED FOR TRAINING}
\label{appendix:features}

The input feature vector \( \mathbf{x} \) used for training both the TBKAN and TBNN models comprises two primary categories: heuristic scalar features derived from the $k$–$\omega$ SST baseline RANS simulations and scalar invariants constructed from the various tensor combinations that appear in the integrity basis used in the expansion of the anisotropy tensor. These features were selected based on their Galilean and rotational invariance, their physical relevance to turbulence in the flows studied, and to the additional constraint that the features are not identically zero owing to the dimensional of the mean flow and/or the presence of a symmetry in the mean flow.

\subsection{\texorpdfstring{Heuristic features ($q_1$ to $q_6$)}{Heuristic features (q1 to q6)}}

These quantities are directly computable from the baseline RANS fields and are commonly used in traditional eddy-viscosity models:

\begin{itemize}
    \item \( q_1 = \min\left( \sqrt{\frac{k y_w}{50\nu}}, 2 \right) \) \\
    Wall-distance-based Reynolds number; clipped to remain finite near walls;
    
    \item \( q_2 = \frac{k}{\epsilon} \sqrt{2 S_{ij}S_{ij}} \) \\
    Ratio of turbulent to mean strain timescales;
    
    \item \( q_3 = \frac{\sqrt{\tau_{ij} \tau_{ij}}}{k} \) \\
    Ratio of Reynolds stress magnitude to turbulent kinetic energy;
    
    \item \( q_4 = \frac{\sqrt{k}}{0.09 \omega y_w} \) \\
    $k$-$\omega$ SST blending function 1 used for transition detection;
    
    \item \( q_5 = \frac{500 \nu}{y_w^2 \omega} \) \\
    $k$-$\omega$ SST blending function 2 sensitive to near-wall gradients;
    
    \item \( q_6 = \min\left( \max(q_4, q_5), \frac{2k}{\max\left( \frac{y_w^2}{\omega}, \nabla k \cdot \nabla \omega \right)} \right) \) \\
    Composite $k$-$\omega$ SST blending feature encoding anisotropy and dissipation scale effects.
\end{itemize}

\subsection{\texorpdfstring{Primary invariant features (\( \lambda_i \))}{Primary invariant features (lambda\_i)}}

A filtered subset of scalar invariants was constructed from a minimal integrity basis involving combinations of normalized strain-rate tensor \( \hat{S}_{ij} \), rotation-rate tensor \( \hat{R}_{ij} \), and antisymmetric tensors \( \hat{C}_{ij} \) and \( \hat{D}_{ij} \) derived from spatial gradients of \( k \) and \( \omega \), respectively. Invariants used in this work include the first (trace) and second  (quadratic form) invariants of the selected tensors:

\begin{itemize}
    \item \( I_1(\mathbf{A}) = \text{Tr}(\mathbf{A}) \), \( I_2(\mathbf{A}) = \frac{1}{2} \left( (\text{Tr}(\mathbf{A}))^2 - \text{Tr}(\mathbf{A}^2) \right) \)
    \item Selected tensors include \( B_{ij}^{(1)} = \hat{S}_{ik} \hat{S}_{kj} \), \( B_{ij}^{(3)} = \hat{R}_{ik} \hat{R}_{kj} \), \( B_{ij}^{(4)} = \hat{C}_{ik} \hat{C}_{kj} \), \( B_{ij}^{(6)} = \hat{R}_{ik} \hat{R}_{kl} \hat{S}_{lj} \), and others as detailed in Table 4 of~\citet{mcconkey2024realizability}
    \item Only invariants that are non-zero for the square duct and periodic hills flow configurations (as determined using symbolic computation based on the computer algebra system SymPy~\cite{Meurer2017}) were retained. The features for the flows studied herein have been archived by~\citet{mcconkey2022deep} and is available from the url~\cite{mcconkey2024optimal_tensor_basis}
\end{itemize}

The primary invariants proposed by \citet{Wu2018} have been post-processed to remove any quantities that are identically zero due to the dimensionality and/or symmetry of the mean flow. As a result, a total of 47 non-trivial scalar invariants were retained and used for training either the TBNN and TBKAN models for the square duct and periodic hills flows.

\subsection{Other scalar features}
\begin{itemize}
    \item Wall-distance-based Reynolds number: \( Re_y = \frac{y_w \sqrt{k}}{\nu} \)
    \item Turbulence intensity ratio: \( \frac{k}{\epsilon + \omega} \)
\end{itemize}

All input features were standardized by removing their mean and normalizing by their standard deviation prior to their use in network training.

\section*{References}
\bibliography{aipsamp}
\end{document}